
\input eplain

\newcount\fignumber
\def\figdef#1{\global\advance\fignumber by 1 \definexref{#1}{\number\fignumber}{figure}\ref{#1}}
\def\figdefn#1{\global\advance\fignumber by 1 \definexref{#1}{\number\fignumber}{figure}}
\let\figref=\ref
\let\figrefn=\refn
\let\figrefs=\refs

%
\ifx\pdfoutput\undefined
\input epsf
\def\fig#1{\epsfbox{#1.eps}}

%
\else
\def\fig#1{\pdfximage {#1.pdf}\pdfrefximage\pdflastximage}

\fi

\newcount\scount \scount=0
\def\section#1{\global\advance\scount by1
    \vskip.25truein\noindent\the\scount\quad{\bf #1}\hfill\vskip.25truein}

\centerline{\bf{Recurrence for Pandimensional Space-Filling Functions}}
\bigskip
\centerline{Aubrey Jaffer\numberedfootnote{Present Address: Digilant, 100 North Washington Street Suite 502, Boston, MA 02114.}}
\medskip
\centerline{{\tt agj@alum.mit.edu}}
\bigskip

\beginsection{Abstract}

{\narrower

A space-filling function is a bijection from the unit line segment to
the unit square, cube, or hypercube.  The function from the unit line
segment is continuous.  The inverse function, while well-defined, is
not continuous.  Space-filling curves, the finite approximations to
space-filling functions, have found application in global
optimization, database indexing, and dimension reduction among others.
For these applications the desired transforms are mapping a scalar to
multidimensional coordinates and mapping multidimensional coordinates
to a scalar.

Presented are recurrences which produce space-filling functions and
curves of any rank $d\ge2$ based on serpentine Hamiltonian paths on
$({\bf Z}\bmod s)^d$ where $s\ge2$.  The recurrences for inverse
space-filling functions are also presented.  Both Peano and Hilbert
curves and functions and their generalizations to higher dimensions
are produced by these recurrences.  The computations of these
space-filling functions and their inverse functions are absolutely
convergent geometric series.

The space-filling functions are constructed as limits of integer
recurrences and equivalently as non-terminating real recurrences.
Scaling relations are given which enable the space-filling functions
and curves and their inverses to extend beyond the unit area or volume
and even to all of $d$-space.

This unification of pandimensional space-filling curves facilitates
quantitative comparison of curves generated from different Hamiltonian
paths.  The isotropy and performance in dimension reduction of a
variety of space-filling curves are analyzed.

For dimension reduction it is found that Hilbert curves perform
somewhat better than Peano curves and their isotropic variants.

\par}

\beginsection{Keywords}

{\narrower space-filling functions; space-filling curves; dimension reduction\par}

\beginsection{Table of Contents}

\item{1} Introduction
\itemitem{1.1} Relation to Previous Work
\itemitem{1.2} Symbols
\item{2} Scaling
\item{3} Unit Cells
\itemitem{3.1} Diagonal Travel
\itemitem{3.2} Adjacent Travel
\item{4} Sub-Cell Orientation
\itemitem{4.1} Adjacent-Corner Cells
\item{5} Aligning Sub-Cells
\item{6} The Recurrence
\item{7} The Inverse Recurrence
\item{8} Non-Terminating Recurrence
\item{9} Isotropy
\itemitem{9.1} Adjacent-Corner Isotropy
\item{10} Filling All of Space
\item{11} Dimension Reduction Performance
\item{12} Praxis
\item{13} Conclusion
\item{} References

\beginsection{1 Introduction}

A space-filling function $F(y): [0,1)\to [0,1)^d$ is a continuous
bijection (but not a homeomorphism) from the unit line segment to the
unit square, cube, or hypercube.  Demonstrating the existence of a
one-to-one mapping between points in the line and the plane was the
original motivation for space-filling functions.  It being possible to
construct $F(y)$ as a continuous function is remarkable and results in
there being practical applications for space-filling functions.

There being uncountably many space-filling functions, actual
computation requires a succinct formulation.  Restriction to
self-similar functions is a step towards this goal.  The use of a
finite geometric kernel (the cell) allows separation of the recurrence
from the kernel, which enables quantitative comparisons of the
performance of different cells (with the recurrence) for uses such as
dimension reduction.

The recurrence $Q(u)$ is defined in terms of an integer input and
vector of integers output.  As the recurrence is scaled it approaches
$F(y)$.  At finite scaling factors the image of the recurrence is a
uniform grid of points in the $d$-dimensional cube.  The space-filling
curve $C(y)$ interpolates unit line segments between $Q\left(\lfloor
y\rfloor\right)$ and $Q\left(\lceil y\rceil\right)$, yielding the
familiar winding line representations.

\beginsection{1.1 Relation to Previous Work}

While there are many papers giving algorithms for space-filling curves
in 2 dimensions, there are only a few addressing space-filling curves
of any rank ($d\ge2$).

Butz \cite{Butz1968314}, \cite{Butz1969128}, \cite{10.1109/T-C.1971.223258}
gives (different) algorithms for multidimensional Hilbert and Peano
curves.  Lawder \cite{Lawder00calculationof} makes some improvements
to Butz's Hilbert algorithm.

Alber and Niedermeier \cite{DBLP:conf/cocoon/AlberN98} investigate
Hilbert curves generated by a recurrence and Manhattan Hamiltonian
path cells they term generators.
They suggest that their method might be extensible to side-lengths
greater than 2, but their approach is not taken here.  The result of a
de-novo analysis, the recurrences and serpentine generator presented
here work for all ranks and side-lengths greater than 1.

Jin and Mellor-Crummey \cite{Jin:2005:SFE:1055531.1055537} use
recursion with a table of enumerated states to compute the
space-filling inverse function.  In order to create the Hilbert cell
table, they recommend using the Butz algorithm.  The serpentine
generator presented here creates working cells for all ranks and
side-lengths greater than 1.

The precession of axes (developed here), which improves isotropy for
diagonally opposite corner cells, works with cells having $s^d$ nodes,
where $d\ge2$ is the rank and $s\ge2$ is the side length.  In order to
implement the precession of axes, Jin and Mellor-Crummey cells would
require $s^{d^2}$ nodes.

Moon et al \cite{Moon:2001} find that Hilbert curves are
asymptotically isotropic, as is also determined by the analysis here.
Presented here are pandimensional variants of Peano curves which are
asymptotically isotropic (traditional Peano curves are not).

All of the algorithms cited so far require that the integer precision
for conversion calculations be fixed.  In contrast, the $Q(u)$ and
$q(V)$ functions developed here automatically scale the precision by
powers of $s^d$, yielding the same continuous curves as would be
produced by fixed precision conversions (ignoring reflection and
rotation).

Sagan \cite{Sagan:1994} presents several methods for mapping from the
unit real interval to unit squares or cubes, but not for mapping from
the unit square or cube to the unit line.  His {\it arithmetization}
process transforms coordinates using {\it similarity transforms} which
are constructed through examination of the curve.  In contrast, my
serpentine algorithm generates alignment transforms for space-filling
curves of any rank $d\ge2$ and side-length $s\ge2$.

Sagan's figures 3.71 and 3.72, {\it Peano curves of the switch-back
type} (which he attributes to Walter Wunderlich) match the Peano and
nearly isotropic Peano curves of \figref{PeanoIso} here, but not my
isotropic variant.  His figure 3.73, Wunderlich's {\it Peano curve of
the meander type} corresponds to the $s=3,d=2$ curve
of \figref{corner} of this paper.

\beginsection{1.2 Symbols}

Scalar variables and functions are named by lowercase letters.
Vector-valued variables and functions are named by uppercase letters.
A (zero-based) subscript selects one coordinate of a vector-valued
variable or function.
\smallskip
{\narrower
\itemitem{} {\bf Variables}:

\itemitem{$d\ge 2$} integer rank of the image;
\itemitem{$s\ge 2$} integer cell width;
\itemitem{$0\le t<s^d$} integer Hamiltonian index;
\itemitem{$0\le i<d$} integer index of cell advance dimension (constant for each Hamiltonian-path);
\itemitem{$0\le j<d$} integer index of dimension;
\itemitem{$0\le n$} integer;
\itemitem{$u$} integer;
\smallskip
\itemitem{} {\bf Functions}:

\itemitem{$H:$}$({\bf Z}\bmod s^d)\to({\bf Z}\bmod s)^d$ Hamiltonian-path;
\itemitem{$h:$}$({\bf Z}\bmod s)^d\to({\bf Z}\bmod s^d)$ Hamiltonian-path inverse;
\itemitem{$k:$}$({\bf Z}\bmod s^d)\to({\bf Z}\bmod d)$ index of sub-cell advance dimension;
\itemitem{$\sigma:$}$({\bf Z}\bmod s^d)\to({\bf Z}\bmod d)$ dimension rotation count;
\itemitem{$N:$}$({\bf Z}\bmod s^d)\to({\bf Z}\bmod 2)^d$ sub-cell entry orientation;
\itemitem{$X:$}$({\bf Z}\bmod s^d)\to({\bf Z}\bmod 2)^d$ sub-cell exit orientation;
\itemitem{$A(V,t,w)$} $\to({\bf Z}\bmod 2)^d$ sub-cell alignment;
\itemitem{$A^{-1}(V,t,w)$} $\to({\bf Z}\bmod 2)^d$ sub-cell alignment inverse;
\itemitem{$Q:$}${\bf N}\to{\bf N}^d$ integer sequence;
\itemitem{$q:$}${\bf N}^d\to{\bf N}$ integer sequence inverse;
\itemitem{$C:$}$[0,\infty)\to[0,\infty)^d$ curve;
\itemitem{$F:$}$[0,1)\to[0,1)^d$ space-filling function;
\itemitem{$f:$}$[0,1)^d\to[0,1)$ space-filling function inverse;
\itemitem{$E:$}$[0,1)\to[0,1)^d$ space-filling function;
\itemitem{$e:$}$[0,1)^d\to[0,1)$ space-filling function inverse.
\par}
\smallskip
Both $A$ and $A^{-1}$ are vector-valued.
The term ``rank'' is the number of dimensions $d$.

\beginsection{2 Scaling}

It being more convenient to work with integers, the customary practice
is to define $F(y)$ as the limit of a scaled self-similar integer
injection:

$$\eqalignno{
\left\lfloor{Q\left(u s^{d}\right)\big/s}\right\rfloor&=Q(u)&\eqdef{selfsim}\cr
F(y)&=\lim_{n\to\infty}Q\left(\left\lfloor ys^{dn}\right\rfloor\right)/s^{n},\qquad0\le y<1
&\eqdef{Flim}\cr
}$$

Considering $u$ to be composed of digits in base-$s^d$ and the
coordinates returned as digits in base $s$, equation \eqref{selfsim}
constrains $Q(u)$ so that the higher order digits of the coordinates
it returns are independent of lower-order digits in $u$.

In pursuit of self-similarity, the equation \eqref{Flim} can be
strengthened by linking $F$ and $Q$ for any (integer) $n>0$, $s\ge2$,
(rank) $d\ge2$ and $0\le u<s^{dn}$.

$$Q(u) = \left\lfloor s^{n} F\left(u/s^{dn}\right)\right\rfloor \eqdef{Qlink}$$

Equation \eqref{Qlink} holds for unit cells where the travel direction
of the first node (sub-cell) matches the net travel direction of the
cell; for example, Peano, but not Hilbert cells.  The more general
case which will be used for the rest of the analysis is:

$$\eqalignno{
Q(u) &= \left\lfloor{Q\left(us^{d^2}\right)\big/s^d}\right\rfloor&\eqdef{Qself2}\cr
F(y) &= \lim_{n\to\infty}Q\left(\left\lfloor ys^{d^2n}\right\rfloor\right)\big/s^{dn},\qquad0\le y<1 &\eqdef{Flim2}\cr
Q(u) &= \left\lfloor s^{dn} F\left(u/s^{d^2n}\right)\right\rfloor, \qquad 0 \le u < s^{d^2n} &\eqdef{Qlink2}\cr
}$$

This groups $d$ digits of base $s^d$ into one digit of base $s^{d^2}$.
Although for some unit cells fewer digits could be grouped, grouping
$d$ digits allows all orientations of the unit cell to be cycled
through, improving the isotropy of the resulting curve.

By equation \eqref{Flim2}:

$$F(0)=(0,\dots,0)\eqdef{F(0)}$$

By equations \eqref{F(0)} and \eqref{Qlink2}:

$$Q(0)=(0,\dots,0)\eqdef{Q(0)}$$

A novel approach (for pandimensional space-filling functions)
presented here is the non-terminating recurrence $E(y)$, for which
more elegant scaling holds:

$$\eqalignno{
    E\left({y\over s^{d^2}}\right)
 &={E(y)\over s^d}\qquad 0\le y<1 &\eqdef{Eself}\cr
    E(y) &= F(y)\qquad 0\le y<1 &\eqdef{E=F}\cr
} $$

\beginsection{3 Unit Cells}

The unit cell is defined on a cubic grid of $s^d$ points as a
Manhattan Hamiltonian-path $H(t)$ of coordinate vectors in $({\bf
Z}\bmod s)^d$.  This path visits each grid point exactly once using
only unit-length axis-parallel segments.

All unit cells have corner points; when $s=2$ all of the points are
corners.  This analysis will concern cells whose entry and exit points
are corners.  Without loss of generality, the entry point for every
cell will have coordinates $(0,\dots,0)$.

Of particular interest are what here are termed serpentine paths,
which create unit cells for any rank $d\ge2$ and side-length $s\ge2$.
Serpentine paths have full-length runs parallel to one axis and
unit-length runs in the other directions.  For serpentine cells, the
exit and entry points are either adjacent corners (\figref{adj}) or
furthest diagonally opposite corners (\figref{diag}).  This
investigation is concerned with adjacent (even or odd $s$) and
opposite (odd $s$) cell corners.  Adjacent corner paths with odd $s$
exist (\figref{corner}), but are not serpentine.

\smallskip\verbatim|
(define (make-serpentine-path rank side)
  (define (mspr path rnk)
    (define (msps seq cn)
      (if (negative? cn)
          seq
          (msps (append (map (lambda (coords) (cons cn coords))
                             (if (odd? cn) (reverse path) path))
                        seq)
                (- cn 1))))
    (if (>= rnk rank)
        path
        (mspr (msps '() (- side 1))
              (+ rnk 1))))
  (mspr '(()) 0))
|endverbatim
\centerline{\figdef{make-serpentine-path}}
\smallskip
The Scheme procedure in \figref{make-serpentine-path} returns a list
of the (list of integer) coordinates of a serpentine path having the
given rank and side-length.  In internal procedure {\tt mspr}, {\tt path} is
the list of coordinate lists for a cell of rank ${\tt rnk}-1$.  The
internal procedure {\tt msps} appends {\tt side} copies (alternating reversal)
of {\tt path} with copy-number {\tt cn} adjoined to the front of each
coordinate list.

\beginsection{3.1 Diagonal Travel}

For the diagonal case, a serpentine pattern can fill a unit cell when
$s$ is odd (\figref{diag}).
\smallskip
\vbox{\settabs 3\columns
\+\hfill\fig{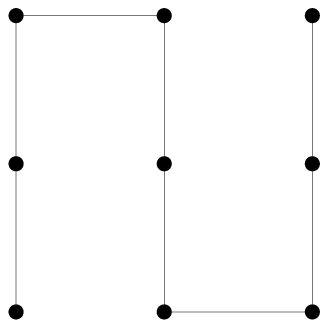}\hfill&
  \hfill\fig{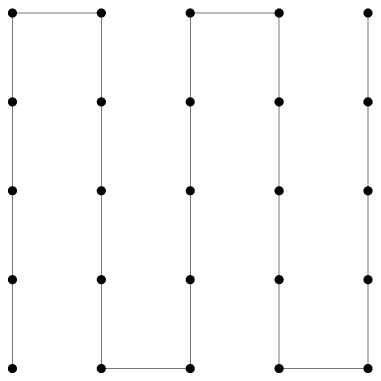}\hfill&
  \hfill\fig{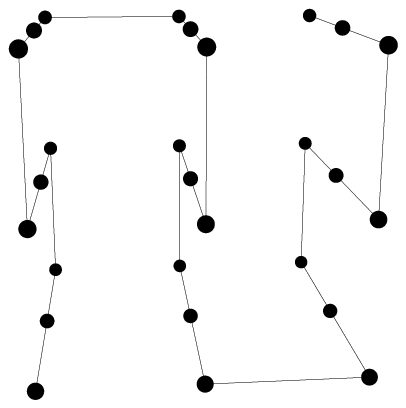}\hfill&\cr
\+\hfill$s=3, d=2$\hfill&
  \hfill$s=5, d=2$\hfill&
  \hfill$s=3, d=3$\hfill&\cr}
\centerline{\figdef{diag}}

For the case of $s$ is even, Dr. Glenn Iba points out that if the
points are alternately labeled black and white, then the starting and
ending points must be opposite colors, which happens only when the
rank $d$ is odd.  Even with odd $d$, the path reaches the outside
corner of each corner sub-cell, from which it can't reach any other.
In order to reach each node in the cell, each straight run must have
odd length.  Thus for use with these recurrences, diagonally opposite
corner cells can be constructed only when $s>2$ is odd.

\beginsection{3.2 Adjacent Travel}

For adjacent entry and exit corners when $s$ is even, a serpentine
pattern fills the square (\figref{adj}).  Stacking an even number of
those squares and connecting the exits and entries results in adjacent
entry and exit corners.  This generalizes to all $d\ge 2$.
\smallskip
\vbox{\settabs 3\columns
\+\hfill\fig{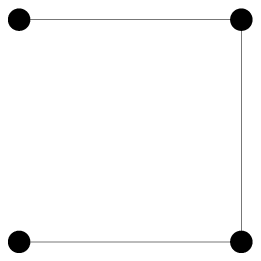}\hfill&
  \hfill\fig{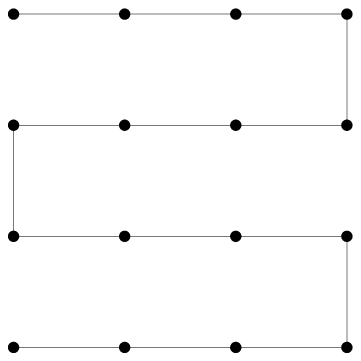}\hfill&
  \hfill\fig{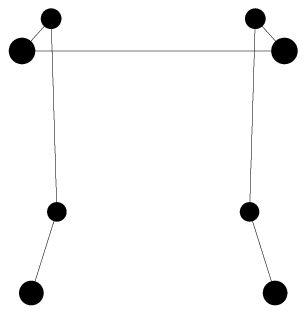}\hfill&\cr
\+\hfill$s=2, d=2$\hfill&
  \hfill$s=4, d=2$\hfill&
  \hfill$s=2, d=3$\hfill&\cr}
\centerline{\figdef{adj}}
\smallskip
When $s$ is odd, a concentric pattern like that shown
in \figref{corner} fills the square and has adjacent entry and exit
corners.  Stacking an odd number of these squares and connecting them
vertically results in entry and exit points which are not adjacent.
But stacking $d-1$ cubes of rank $d-1$ which are rotated $90^{\circ}$
to each other results in a path between diagonally opposite corners of
the $d$-cube.  Stacking an additional cube of rank $d-1$ with a
diagonal path brings the entry and exit points to be above one
another, resulting in the entry and exit points being vertically
adjacent.  Height can be increased by adding pairs of oppositely
aligned diagonal cubes of rank $d-1$.
\smallskip
\vbox{\settabs 3\columns
\+\hfill\fig{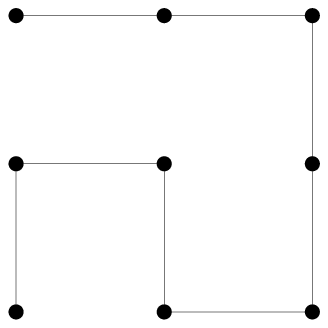}\hfill&
  \hfill\fig{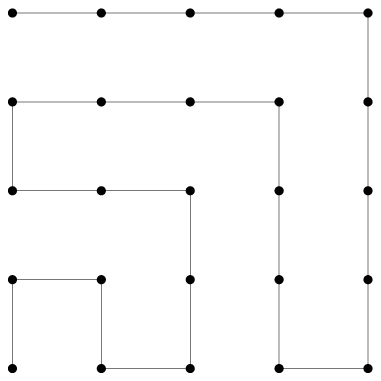}\hfill&
  \hfill\fig{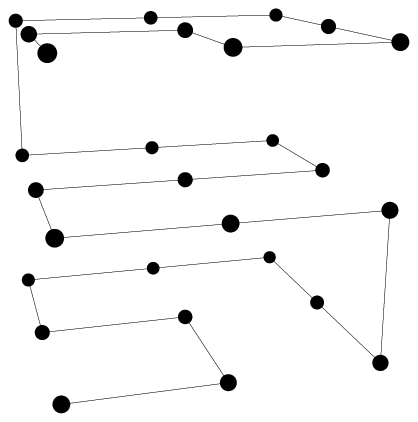}\hfill&\cr
\+\hfill$s=3, d=2$\hfill&
  \hfill$s=5, d=2$\hfill&
  \hfill$s=3, d=3$\hfill&\cr}
\centerline{\figdef{corner}}
\smallskip
So adjacent-corner cells can be constructed when $s\ge 2$ is even and
when $s$ is odd and $s\ge d\ge 2$.

\beginsection{4 Sub-Cell Orientation}

The recurrence to be developed replaces each node of the parent cell
with a (possibly reflected, possibly rotated) child cell whose edge
length is the edge length of its parent cell divided by $s$.  Each
edge of the parent cell is replaced with a short edge (parent edge
length divided by $s$) parallel to the edge being replaced.  The child
cell has $s^d$ points connected by $s^d-1$ edges.

The challenge is to systematically select the orientation of the child
cell so that its entry and exit points connect to its immediate
neighbor cells with single short axis-parallel segments.

Given the $H(t)$ sequence of vectors, define vector sequences $N(t)$
and $X(t)$ to describe the entry and exit orientations.  Because the
entry and exit points are always on corners of the sub-cell, the
coordinates in each vector need only take two values.  Thus vectors
$N(t)$ and $X(t)$ can be elements of $({\bf Z}\bmod 2)^d$.

Because $H(0)=(0,\dots,0)$, it follows that $N(0)=(0,\dots,0)$.
Similarly $(s-1)\cdot X(s^d-1)=H(s^d-1)$.  The relation between the
entrance of a cell and the exit of its predecessor is (with coordinate
subscript $0\le j<d$):

$$\eqalign{
N_j(t) &= \cases
          {0,                                & if $t = 0$;\cr
          X_j(t-1) + H_j(t)-H_j(t-1) \bmod 2, & otherwise.}\cr} \eqdef{N_j}$$

For diagonal-travel cells $X(s^d-1)=(1,\dots,1)$.  For adjacent-travel
cells $X(s^d-1)$ differs from $(0,\dots,0)$ by having $s-1$ in only
one coordinate.

For diagonal-travel $H$-paths, $N(t)$ and $X(t)$ must have opposite
parity in all coordinates:

$$X_j(t) = 1 - N_j(t) \eqdef{X_j}$$

For adjacent-travel $H$-paths the mapping from $N(t)$ to $X(t)$ is
more complicated.

Consider the first sub-cell, which replaces the $H(0)$ node.  In order
to reach the next sub-cell, the first sub-cell must traverse a
distance of $s$ along one of its axes.  The only way to do this is for
$N(1)-X(0)$ (traversing $s-1$) to be in the same direction as
$H(1)-H(0)$, which will be non-zero in only one coordinate.
Similarly, the cell which replaces the $H(s^d-1)$ node must align
$X(s^d-1)-N(s^d-1)$ with $H(s^d-1)-H(s^d-2)$.

\smallskip
\vbox{\centerline{\fig{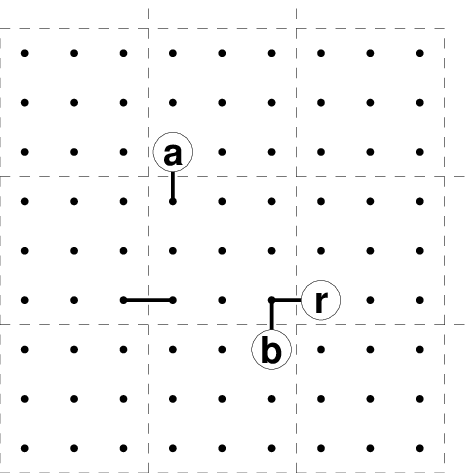}}
      \centerline{\figdef{NX3cell2}}}

Now consider a $d=2$ cell at the center of \figref{NX3cell2} which is
neither first nor last, indexed by $0<t<s^d-1$.  $N(t)$ is a corner of
the sub-cell.  Without loss of generality assume that $N(t)$ is
$(0,0)$ and that the previous cell is to the left of the current cell.
After visiting all the points in the current cell there are 3 cells
that the path could connect to: right({\bf r}), above({\bf a}), and
below({\bf b}).

\item{({\bf r})} If $H(t+1)-H(t)=(1,0)$, then $X(t)=(1,0)$ and net travel is to the
right.

\item{({\bf a})} If $H(t+1)-H(t)=(0,1)$, then $X(t)=(0,1)$ and connects to the
bottom left corner of the cell above.

\item{({\bf b})} If $H(t+1)-H(t)=(0,-1)$, then $X(t)=(1,0)$ and connects to the top
right corner of the cell below.

In cases ({\bf r}) and ({\bf a}), $X(t)-N(t)$ must be in the same
direction as $H(t+1)-H(t)$ in order to reach the next cell.  In case
({\bf b}) $X(t)$ is already adjacent to the cell below, so $X(t)-N(t)$
can't be in the direction of $H(t+1)-H(t)$.  When $d>2$ there are more
possible directions for $X(t)-N(t)$ which may or may not be allowed
depending on where $H(t)$ is in the parent cell.  The direction which
is always allowed is $H(t)-H(t-1)$, so choose it.\numberedfootnote{The
note at the end of the next section discusses an exception.}

\beginsection{4.1 Adjacent-Corner Cells}

Generalizing this to $d\ge 2$ and all orientations, $H(t+1)-H(t)$ will
be non-zero in only one coordinate; for $0<t<s^d-1$, let $0\le k(t)<d$
be the index of the non-zero coordinate in $H(t+1)-H(t)$.

The first and last sub-cells must terminate at adjacent corners of the
cell.  Because $H(0)=(0,\dots,0)$, it follows that $N(0)=(0,\dots,0)$
and $X(0)=H(1)-H(0)=H(1)$.

$$X_j(s^d-1)={H_j(s^d-1)-H_j(0)\over s-1}={H_j(s^d-1)\over s-1} \eqdef{X_jl}$$

The rest of the cases concern $0<t<s^d-1$.

If $N_{k(t)}(t)=0$ and $H_{k(t)}(t+1)-H_{k(t)}(t)=1$ or
$N_{k(t)}(t)=1$ and $H_{k(t)}(t+1)-H_{k(t)}(t)=-1$, then the net
travel through the cell $X(t)-N(t)$ will be in the same direction as
$H(t+1)-H(t)$:

$$X_j(t) = \cases
          {1 - N_j(t),                & if $j = k(t)$;\cr
          N_j(t),                     & otherwise.\cr} \eqdef{X_j1}$$

If that is not the case, then travel through the cell will be in the
direction of $H(t)-H(t-1)$:

$$X_j(t) = \cases
          {1 - N_j(t),                & if $j = k(t-1)$;\cr
          N_j(t),                     & otherwise.\cr} \eqdef{X_j2}$$

The cases of $X_j(0)$ and $X_j(s^d-1)$ can be absorbed by formulas \eqref{X_j1}
and \eqref{X_j2} by defining $k(0)$ to have the same value as $k(1)$, and
$k(s^d-1)$ to have the same value as $k(s^d-2)$.
\medskip
{\bf Note:} While computer tests show that these rules work with all
serpentine cells of up to ($s^d\le$) $10^6$ points, they do not work
with every adjacent-corner Manhattan Hamiltonian-path on $({\bf
Z}\bmod s)^d$ with odd~$s$.  The right side of \figref{bad} shows a
non-serpentine cell for which equations \eqref{X_j1} and \eqref{X_j2}
don't work.  It is unknown whether there exists rules which will work
with that cell.
\medskip
\vbox{\settabs 2\columns
\+\hfill\fig{adj3cell3}\hfill&\hfill\fig{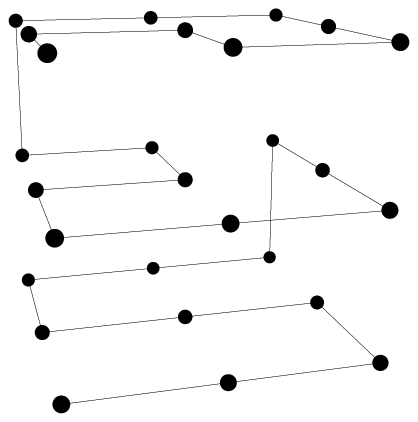}\hfill&\cr
\+\hfill$s=3,d=3$ adjacent-corners\hfill&\hfill$s=3,d=3$ adjacent-corners\hfill&\cr
\+&\hfill problem for formulas \eqref{X_j1}, \eqref{X_j2}\hfill&\cr}
\centerline{\figdef{bad}}

\beginsection{5 Aligning Sub-Cells}

The integer recurrence will use $s^n H(t)$ as the coordinates of the
parent layer; each sub-cell (having coordinates $s^{n-1} H(t')$) will
be reflected and rotated to match orientation with $N(t)$ and $X(t)$.

For diagonal travel unit-cells, all coordinates change between $N(t)$
and $X(t)$; so once the sub-cell is aligned with $N(t)$, it will also
be aligned with $X(t)$.  The sub-cell can be aligned with $N(t)$ by
simply complementing each coordinate in the sub-cell for which the
corresponding coordinate in $N(t)$ is nonzero:

$$A_j(V,t,w) = \cases
              {V_j,                  & if $N_j(t) = 0$;\cr
              w-V_j,                 & otherwise.\cr} \eqdef{A_j}$$

In adjacent travel cells, $H(s^d-1)$ has only one nonzero coordinate.
Because the recurrence is self-similar, for each $t$, $N(t)-X(t)$ also
has only one nonzero coordinate.  When $N(t)-X(t)$ and $H(s^d-1)$ are
nonzero in the same coordinate position, then $A_j(V,N,s)$ defined in
equation \eqref{A_j} will align the sub-cell with $N(t)$ and $X(t)$.

If $N(t)-X(t)$ and $H(s^d-1)$ are nonzero in different coordinate
positions $0\le k(t)<d$ and $0\le i<d$ respectively, then the $i$th
coordinate of the sub-cell must be moved to the $k(t)$th coordinate
position.  This can be done by applying a cyclic permutation to the
coordinates $V$:

$$A_j(V,t,w) = \cases
              {V_{j+i-k(t)\bmod d},      & if $N_j(t) = 0$;\cr
              w-V_{j+i-k(t)\bmod d},     & otherwise.\cr} \eqdef{A_jc}$$

Formula \eqref{A_jc} works also when $k(t)=i$.  Note that
$\sigma(t)=i-k(t)\bmod d$ can be precomputed for each node in the
cell's Hamiltonian-path (see equation \eqref{sigma(t)}).

\beginsection{6 The Recurrence}

$R$ is the recurrence for integer coordinates.  $u$ is the non-negative
integer scalar input; a vector of non-negative integers is returned.
The highest order digits of $u$ are processed first.  At each stage
$R$ orients the coordinates at the next smaller scale $w/s$ so that
the start and end of its Hamiltonian-path matches the
$N(\left\lfloor{u/m}\right\rfloor)$ and
$X(\left\lfloor{u/m}\right\rfloor)$ of the current scale.

$$\eqalign{
l(u) &= \left\lceil\log_{s^{d^2}}(1+u)\right\rceil\cr
R(u,m,w) &= \cases{
            w H\left(\left\lfloor {u/ m}\right\rfloor\right) +
            A\left(R\left({u\bmod m},
              \left\lfloor{m/ s^d}\right\rfloor,
              \left\lfloor{w/ s}\right\rfloor\right),
              \left\lfloor {u/ m}\right\rfloor, w-1\right) & if $w>0$;\cr
         (0,0,\dots) & otherwise.\cr
          }\cr
  Q(u) &= R\left(u, s^{d^2 l(u)-d}, s^{d\cdot l(u)-1}\right)\cr
        } \eqdef{R}$$

$s^{d^2 l(u)}$ is the smallest non-negative integer power of
${s^{d^2}}$ greater than $u$.  $s^{d^2 l(u)-d}$ is one $s^d$ digit
shorter than ${s^{d^2}}$ so that $R$ starts with the most significant
digit instead of a gratuitous digit of zero.  The base for the
logarithm is ${s^{d^2}}$ in order to always work on groups of $d$
(base $s^d$) digits.  This is needed for Hamiltonian-paths where
$X(0)\ne~H(s^d-1)/(s-1)$.

$C(y)$ is the curve which bridges the gaps between successive
(integer) points with unit line segments:

$$C(y) = \left(1+\lfloor y\rfloor-y\right) Q\left(\lfloor y\rfloor\right) +
         \left(y-\lfloor y\rfloor\right) Q\left(\lfloor y\rfloor+1\right)
\eqdef{C(y)}$$

$Q(u)$ can be used as a space-filling function by scaling by $n>0$
groups of digits:

$$F(y) = \lim_{n\to\infty}Q\left(\left\lfloor ys^{d^2n}\right\rfloor\right)\big/s^{dn} \eqno\eqref{Flim2}$$

\figrefs{PeanoF} and \figrefn{HilbertF} show $F(y)$ evaluated at 729 and 1024 points
respectively for 3 values for $n$.  The reason that the $n=3$ curves
don't resemble the $n=1$ curves is because many of the inflection
points of the $n=3$ curve are not being traced.

\smallskip
\vbox{{\settabs 3\columns
\+\hfill\fig{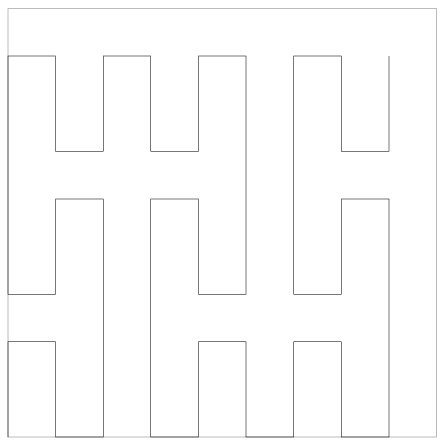}\hfill&
  \hfill\fig{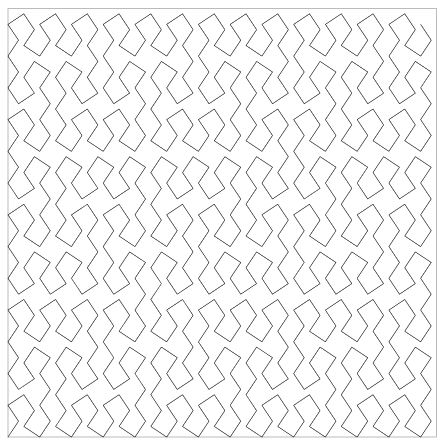}\hfill&
  \hfill\fig{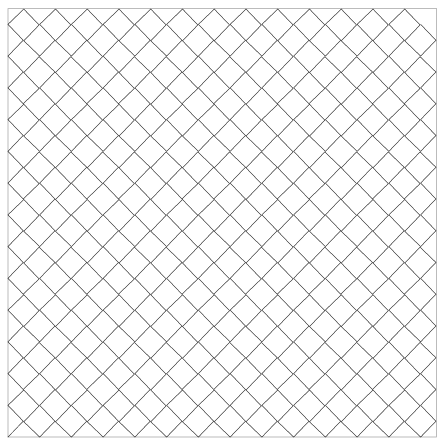}\hfill&\cr
\+\hfill Peano $n=1$\hfill&
  \hfill Peano $n=2$\hfill&
  \hfill Peano $n=3$\hfill&\cr}
\centerline{\figdef{PeanoF}}}

\smallskip
\vbox{{\settabs 3\columns
\+\hfill\fig{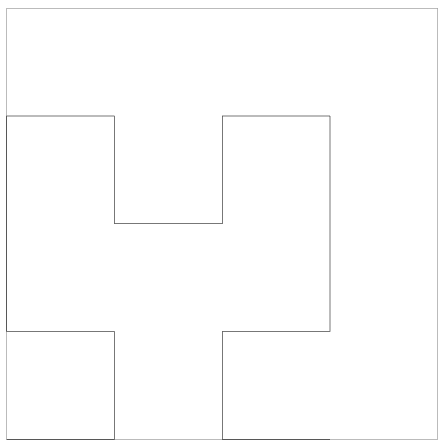}\hfill&
  \hfill\fig{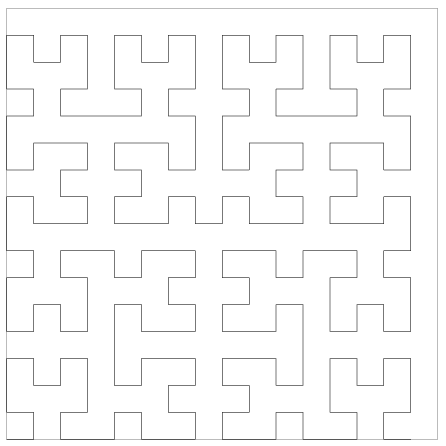}\hfill&
  \hfill\fig{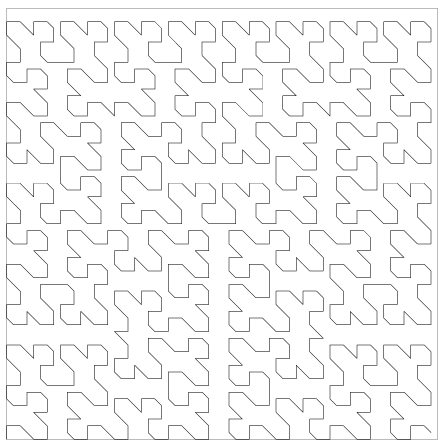}\hfill&\cr
\+\hfill Hilbert $n=1$\hfill&
  \hfill Hilbert $n=2$\hfill&
  \hfill Hilbert $n=3$\hfill&\cr}
\centerline{\figdef{HilbertF}}}
\smallskip

\beginsection{7 The Inverse Recurrence}

Like $Q(u)$, its scalar-valued inverse $q(V)$ processes the highest
order digits of its argument (in this case the coordinates $V$) first.
$s^{d\cdot l(V)}$ is the lowest integer power of $s^d$ greater than
the largest coordinate in $V$.  $s^{d\cdot l(u)-1}$ is one $s$ digit
shorter than ${s^{d}}$ so that $r$ starts with the most significant
digit instead of a gratuitous digit of zero.
$h\left(\left\lfloor{V/w}\right\rfloor\right)$ is the index in $H$ of
the highest base-$s$ digit of the coordinates $V$.

$$\eqalign{
{V\bmod w} &\equiv
\left({V_0\bmod w}, \dots, {V_{d-1}\bmod w}\right)\cr
\left\lfloor{V\over w}\right\rfloor &\equiv
\left(\left\lfloor{V_0\over w}\right\rfloor, \dots, \left\lfloor{V_{d-1}\over w}\right\rfloor\right)\cr
l(V) &= \lceil\log_{s^d}(1+\max(V_0, \dots, V_{d-1}))\rceil\cr
r(u,V,w) &= \cases{
        r\left({us^d+h\left(\left\lfloor{V\over w}\right\rfloor\right)},
                   A^{-1}\left({V\bmod w}, h\left(\left\lfloor{V\over w}\right\rfloor\right), w-1\right),
                    \left\lfloor{w\over s}\right\rfloor
             \right)                    & if $w>0$;\cr
        u                               & otherwise.\cr
          }\cr
  q(V) &= r\left(0, V, s^{d\cdot l(V)-1}\right)\cr
  f(Y) &= \lim_{n\to\infty}q\left(\left\lfloor Ys^{dn}\right\rfloor\right)\big/s^{d^2n}\cr} \eqdef{f}$$

While $A$ as a function of $V$ is its own inverse, $A'$ and $A''$
(introduced in the ``Isotropy'' section) will not be their own
inverses when $d>2$.

$$\eqalignno{
        A'^{-1}_j(V,t,w) &= A_{j-i\bmod d}(V,t,w) &\eqdef{A'^{-1}}\cr
        A''^{-1}_j(V,t,w) &= A_{j-t-1\bmod d}(V,t,w) &\eqdef{A''^{-1}}\cr
        }$$

\beginsection{8 Non-Terminating Recurrence}

Notice that in the alignment function $A(V,t,w)$ nothing constrains
the argument $V$ to be a vector of integers.  It also works as a
vector of real numbers.  By scaling the input by $s^d$ and the output
by $s$ at each call, a non-terminating recurrence (with $0\le y<1)$ can
be written:

$$E(y)=
  { H\left(\left\lfloor s^dy\right\rfloor\right)
   +A\left(E\left(s^dy-\left\lfloor s^dy\right\rfloor\right), \left\lfloor s^dy\right\rfloor, 1\right)
   \over s}\eqdef{E}$$

Because the denominators are increasing powers of $s$, $E(y)$ is an
absolutely convergent geometric series.

$H$, $s$, and $d$ are all characteristics of the cell path $P$.
Writing $A^*$, $H^*$, and $E^*$ as functions of $P$, the similarity to
a fixed-point combinator can be seen:

$$E^*(P,y)=
  { H^*\left(P,\left\lfloor s^dy\right\rfloor\right)
   +A^*\left(P,E^*\left(P,s^dy-\left\lfloor s^dy\right\rfloor\right), \left\lfloor s^dy\right\rfloor, 1\right)
   \over s}\eqdef{E*}$$

Because it doesn't terminate, equation~\eqref{E} can't be computed
directly.  By limiting its depth to $nd$, the curves produced by
equation~\eqref{E''} can be explored:

$$\eqalign{
 E''(y,c)&={1 \over s}
 \cases{
  { H\left(\left\lfloor s^dy\right\rfloor\right)+(1/2,\dots,1/2) },
    & if $c\le1$;\cr
  { H\left(\left\lfloor s^dy\right\rfloor\right)
   +A\left(E''\left(s^dy-\left\lfloor s^dy\right\rfloor, c-1\right), \left\lfloor s^dy\right\rfloor, 1\right) },
   & otherwise.\cr}\cr
 E''(y)&=E''(y,nd)\cr
} \eqdef{E''}$$

\smallskip
\vbox{{\settabs 3\columns
\+\hfill\fig{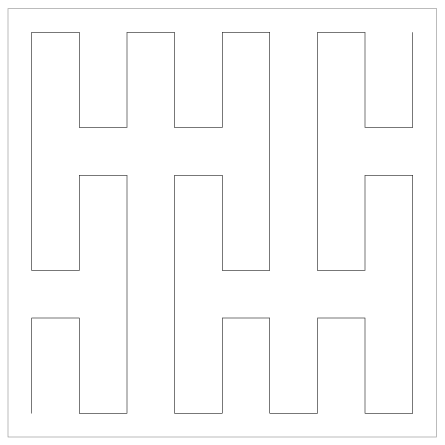}\hfill&
  \hfill\fig{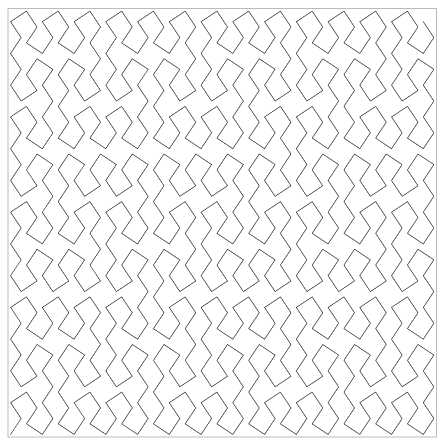}\hfill&
  \hfill\fig{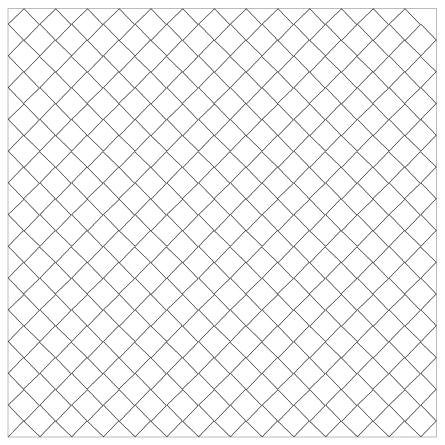}\hfill&\cr
\+\hfill Peano $n=1$\hfill&
  \hfill Peano $n=2$\hfill&
  \hfill Peano $n=3$\hfill&\cr}
\centerline{\figdef{PeanoE}}}

\smallskip
\vbox{{\settabs 3\columns
\+\hfill\fig{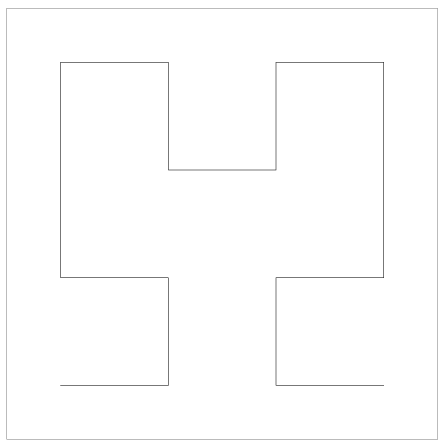}\hfill&
  \hfill\fig{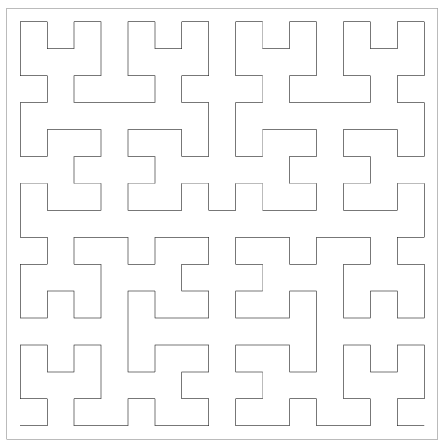}\hfill&
  \hfill\fig{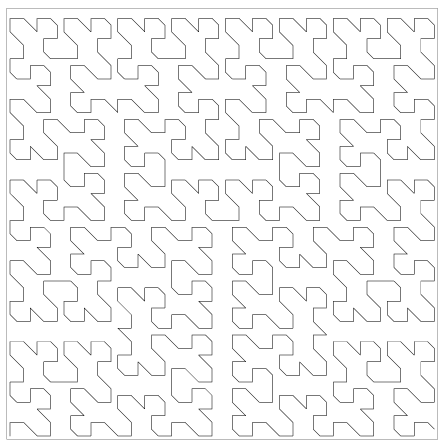}\hfill&\cr
\+\hfill Hilbert $n=1$\hfill&
  \hfill Hilbert $n=2$\hfill&
  \hfill Hilbert $n=3$\hfill&\cr}
\centerline{\figdef{HilbertE}}}
\smallskip

As with \figrefs{PeanoF} and \figrefn{HilbertF}, \figrefs{PeanoE}
and \figrefn{HilbertE} are traced at 729 and 1024 points respectively.
Note that $E''$ terminates at the middle of each $s^{-nd}$ sub-cell;
its plots are centered at $(1/2,\dots,1/2)$ where \figrefs{PeanoF}
and \figrefn{HilbertF} are pinned to the origin.

The inverse non-terminating recurrence is:

$$e(V)=
  { h\left(\left\lfloor sV\right\rfloor\right)
   +e\left(A^{-1}\left(sV-\left\lfloor sV\right\rfloor, h\left(\left\lfloor sV\right\rfloor\right), 1\right)\right)
   \over s^d}\eqdef{e}$$

\beginsection{9 Isotropy}

A complete pattern contains $s^{dn}$ points connected in sequence by
$s^{dn}-1$ edges.  There are $s^{d}$ sub-cells, each of which has
$s^{d(n-1)}-1$ edges.  The total for the sub-cells is $s^{dn}-s^d$
edges; but the edges of the top-level cell connect the sub-cells,
resulting in $s^{dn}-s^d+s^d-1=s^{dn}-1$ edges.

Each edge connecting grid points in a complete pattern is parallel to
one of the axes.  Isotropy is the condition that equal lengths of
edges are parallel to each axis.

The Peano curve on the left in \figref{PeanoIso} has 60 vertical and 20
horizontal segments, much more vertical than horizontal travel.  Each
of its sub-cells could be oriented horizontally or vertically, but all
are vertical.  There are only straight vertical runs of lengths 2 and
5, and all horizontal runs are length 1.

With serpentine diagonal cells, the major axis travel is
$(s-1)s^{d-1}$, the next most traveled axis travel is $(s-1)s^{d-2}$,
etc.  With cells at all levels aligned, the counts will be
proportional, and indeed 60 and 20 are proportional to 6 and 2.

\smallskip
\vbox{\settabs 3\columns
\+\hfill\fig{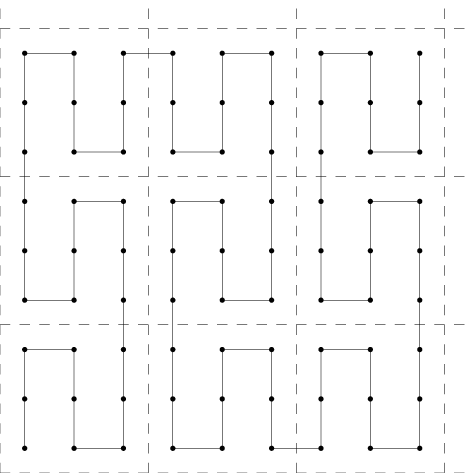}\hfill&
  \hfill\fig{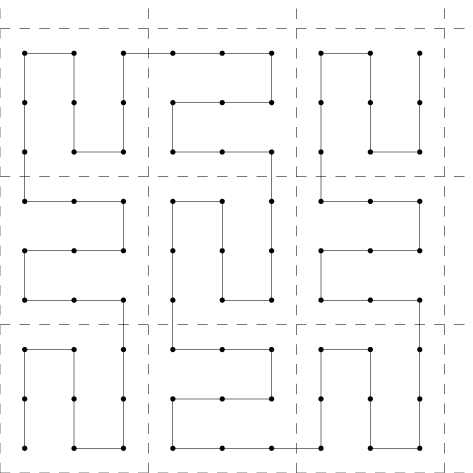}\hfill&
  \hfill\fig{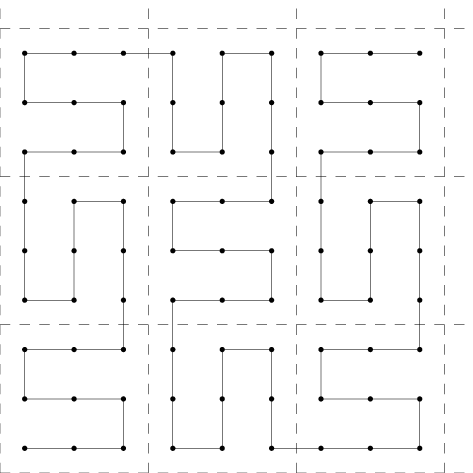}\hfill&\cr
\+\hfill Peano\hfill&
  \hfill Nearly Isotropic\hfill&
  \hfill Isotropic\hfill&\cr}
\centerline{\figdef{PeanoIso}}

If the orientation of the cells is rotated from sub-cell to sub-cell,
then a $s^{d^2}$ point pattern can be constructed which is closer to
isotropic.  This can be done by applying a cyclic permutation to the
coordinates $V$ in the alignment function:

$$A'_j(V,t,w) = \cases
              {V_{j+t\bmod d},              & if $N_j(t) = 0$;\cr
              w-V_{j+t\bmod d},             & otherwise.\cr} \eqdef{A'}$$

The center panel of \figref{PeanoIso} shows the result, having 4
horizontally oriented sub-cells and 5 vertically oriented sub-cells in
the enclosing vertical cell for a total of 44 vertical and 36
horizontal segments.  There are straight runs of length 1, 2, and 3 in
each direction.

If the permutation count is offset by 1 so that the enclosing cell has
a different orientation from the first sub-cell, then balance is
achieved for this $s=3, d=2$ case:

$$A''_j(V,t,w) = \cases
              {V_{j+t+1\bmod d},              & if $N_j(t) = 0$;\cr
              w-V_{j+t+1\bmod d},             & otherwise.\cr} \eqdef{A''}$$

With the opposite sub-cell orientations, but with the same enclosing
cell orientation, the curve on the right side of \figref{PeanoIso} has
40 horizontal and 40 vertical segments.

A serpentine $d$-level ($s^{d^2}$ point) pattern with $A''$ precession
will be isotropic when its $s^d$ sub-cells and 1 parent cell have
orientations evenly distributed among the dimensions.  A necessary
(but not sufficient) condition for isotropy is: $s^{d}+1\equiv0\bmod
d$.  This will not happen when $d=s=3$.  But the precessing $d=s=3$
case is asymptotically isotropic:

$$\vbox{\settabs 9\columns
  \+\hfill nodes&\hfill   &$j=    0$\hfill &\hfill   &$j=    1$\hfill &\hfill   &$j=    2$\hfill &\cr
  \+\hfill    27&\hfill     2&\hfill  7.4\%&\hfill     6&\hfill 22.2\%&\hfill    18&\hfill 66.7\%&\cr
  \+\hfill   729&\hfill   236&\hfill 32.4\%&\hfill   240&\hfill 32.9\%&\hfill   252&\hfill 34.6\%&\cr
  \+\hfill 19683&\hfill  6554&\hfill 33.3\%&\hfill  6558&\hfill 33.3\%&\hfill  6570&\hfill 33.4\%&\cr
}$$

The $A'$ precession gives the same results (just permuted):

$$\vbox{\settabs 9\columns
  \+\hfill nodes&\hfill   &$j=    0$\hfill &\hfill   &$j=    1$\hfill &\hfill   &$j=    2$\hfill &\cr
  \+\hfill    27&\hfill    18&\hfill 66.7\%&\hfill     2&\hfill  7.4\%&\hfill     6&\hfill 22.2\%&\cr
  \+\hfill   729&\hfill   240&\hfill 32.9\%&\hfill   252&\hfill 34.6\%&\hfill   236&\hfill 32.4\%&\cr
  \+\hfill 19683&\hfill  6554&\hfill 33.3\%&\hfill  6558&\hfill 33.3\%&\hfill  6570&\hfill 33.4\%&\cr
}$$

The orientation of a serpentine cell is the direction of its straight
runs of length $s-1$ (all perpendicular runs will have length 1).  In
order for there to be a straight run longer than $s$, two cells and
the edge that connects them must all be aligned.  Precessing the
sub-cell orientations prevents long runs between adjacent sub-cells at
the lowest level.  What about connections between sub-cells of adjacent
parent cells?  Because the cell orientations precess, the shift-count
offset between the exit and entry cells is $1-(s^d-1)\bmod d=-s^d\bmod
d$.  The cells align only when $s^d\equiv0\bmod d$.  While long
straight runs are prevented when $s=3$ and $d=2$, they will occur when
$s=d=3$.

The isotropy of $s=d=3$ can be somewhat improved by reflecting one of
its planes along its diagonal, as shown in \figref{c3diag3ni}:

\smallskip
\vbox{\settabs 1\columns
\+\hfill\fig{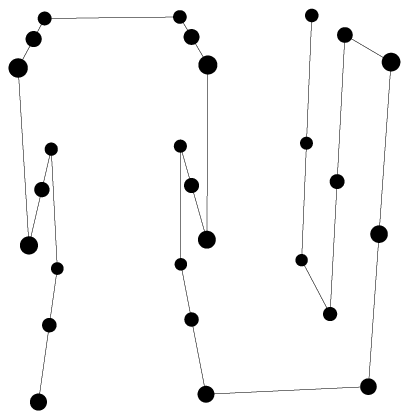}\hfill&\cr
\+\hfill$s=3,d=3$ diagonal-corners\hfill&\cr}
\centerline{\figdef{c3diag3ni}}

This reduces the segment-count disparity between directions, which
results in the complete patterns being closer to evenly divided.

$$\vbox{\settabs 9\columns
  \+\hfill nodes&\hfill   &$j=    0$\hfill &\hfill   &$j=    1$\hfill &\hfill   &$j=    2$\hfill &\cr
  \+\hfill    27&\hfill     2&\hfill  7.4\%&\hfill    10&\hfill 37.0\%&\hfill    14&\hfill 51.9\%&\cr
  \+\hfill   729&\hfill   236&\hfill 32.4\%&\hfill   244&\hfill 33.5\%&\hfill   248&\hfill 34.0\%&\cr
  \+\hfill 19683&\hfill  6554&\hfill 33.3\%&\hfill  6562&\hfill 33.3\%&\hfill  6566&\hfill 33.4\%&\cr
}$$

\beginsection{9.1 Adjacent-Corner Isotropy}

Even though it is asymmetrical, the $s=3, d=2$ cell on the left side
of \figref{corner} is balanced.  All complete $s^{d^2n}$ patterns constructed
with it will be balanced:

$$\vbox{\settabs 9\columns
  \+\hfill nodes&\hfill   &$j=    0$\hfill &\hfill   &$j=    1$\hfill &\cr
  \+\hfill     9&\hfill     4&\hfill 44.4\%&\hfill     4&\hfill 44.4\%&\cr
  \+\hfill    81&\hfill    40&\hfill 49.4\%&\hfill    40&\hfill 49.4\%&\cr
  \+\hfill   729&\hfill   364&\hfill 49.9\%&\hfill   364&\hfill 49.9\%&\cr
  \+\hfill  6561&\hfill  3280&\hfill 50.0\%&\hfill  3280&\hfill 50.0\%&\cr
}$$

With an even $s$, the number of edges in a complete pattern,
$s^{d^2}-1$, will be odd.  The tally of edges by direction for the
2-dimensional Hilbert curve is:

$$\vbox{\settabs 9\columns
  \+\hfill nodes&\hfill   &$j=    0$\hfill &\hfill   &$j=    1$\hfill &\cr
  \+\hfill     4&\hfill     2&\hfill 50.0\%&\hfill     1&\hfill 25.0\%&\cr
  \+\hfill    16&\hfill     7&\hfill 43.8\%&\hfill     8&\hfill 50.0\%&\cr
  \+\hfill    64&\hfill    32&\hfill 50.0\%&\hfill    31&\hfill 48.4\%&\cr
  \+\hfill   256&\hfill   127&\hfill 49.6\%&\hfill   128&\hfill 50.0\%&\cr
}$$

But if one more edge is included in the tally, this particular pattern
achieves perfect balance:

$$\vbox{\settabs 9\columns
\+\hfill nodes&\hfill &$j=    0$\hfill &\hfill &$j=    1$\hfill &\cr
\+\hfill     4&\hfill     2&\hfill 50.0\%&\hfill     2&\hfill 50.0\%&\cr
\+\hfill    16&\hfill     8&\hfill 50.0\%&\hfill     8&\hfill 50.0\%&\cr
\+\hfill    64&\hfill    32&\hfill 50.0\%&\hfill    32&\hfill 50.0\%&\cr
\+\hfill   256&\hfill   128&\hfill 50.0\%&\hfill   128&\hfill 50.0\%&\cr
}$$

The 3-dimensional and 4-dimensional Hilbert cells aren't balanced, but
their patterns are asymptotically balanced:

$$\vbox{\settabs 9\columns
  \+\hfill nodes&\hfill   &$j=    0$\hfill &\hfill   &$j=    1$\hfill &\hfill   &$j=    2$\hfill &\cr
  \+\hfill     8&\hfill     4&\hfill 50.0\%&\hfill     1&\hfill 12.5\%&\hfill     2&\hfill 25.0\%&\cr
  \+\hfill    64&\hfill    18&\hfill 28.1\%&\hfill    22&\hfill 34.4\%&\hfill    23&\hfill 35.9\%&\cr
  \+\hfill   512&\hfill   171&\hfill 33.4\%&\hfill   174&\hfill 34.0\%&\hfill   166&\hfill 32.4\%&\cr
  \+\hfill  4096&\hfill  1374&\hfill 33.5\%&\hfill  1355&\hfill 33.1\%&\hfill  1366&\hfill 33.3\%&\cr
  \+\hfill 32768&\hfill 10902&\hfill 33.3\%&\hfill 10926&\hfill 33.3\%&\hfill 10939&\hfill 33.4\%&\cr
}$$

$$\vbox{\settabs 9\columns
  \+\hfill nodes&\hfill   &$j=    0$\hfill &\hfill   &$j=    1$\hfill &\hfill   &$j=    2$\hfill &\hfill   &$j=    3$\hfill &\cr
  \+\hfill    16&\hfill     8&\hfill 50.0\%&\hfill     1&\hfill  6.2\%&\hfill     2&\hfill 12.5\%&\hfill     4&\hfill 25.0\%&\cr
  \+\hfill   256&\hfill    44&\hfill 17.2\%&\hfill    58&\hfill 22.7\%&\hfill    71&\hfill 27.7\%&\hfill    82&\hfill 32.0\%&\cr
  \+\hfill  4096&\hfill  1002&\hfill 24.5\%&\hfill  1104&\hfill 27.0\%&\hfill  1098&\hfill 26.8\%&\hfill   891&\hfill 21.8\%&\cr
  \+\hfill 65536&\hfill 17011&\hfill 26.0\%&\hfill 16562&\hfill 25.3\%&\hfill 15544&\hfill 23.7\%&\hfill 16418&\hfill 25.1\%&\cr
}$$

For adjacent-corner cells with $d=2$ the orientation of each sub-cell
is constrained by the fact that only one of axes has nonzero
net-travel (there being only two permutations of 2 axes).  For $d>2$
there is wobble possible by using permutations other than powers of
the one cyclic permutation.  This will not make the cell isotropic but
could make the $s^{d^2n}$ patterns become more isotropic as $n$
increases.

\smallskip
\vbox{\settabs 2\columns
\+\hfill\fig{adj3cell3}\hfill&
  \hfill\fig{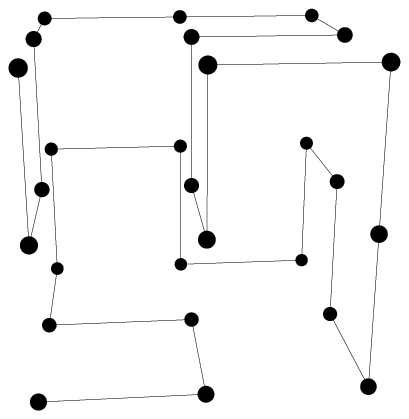}\hfill&\cr
\+\hfill$s=3,d=3$ adjacent-corners\hfill&
  \hfill$s=3,d=3$ adjacent-corners\hfill&\cr}
\centerline{\figdef{adjcorn}}

Because the start and end segments are not part of straight runs, the
maximum run length for curves formed from the cell on the left side
of \figref{adjcorn} (the same as the cell from \figref{corner}) is 3.
Despite that, small patterns from that cell are quite anisotropic:

$$\vbox{\settabs 9\columns
  \+\hfill nodes&\hfill   &$j=    0$\hfill &\hfill   &$j=    1$\hfill &\hfill   &$j=    2$\hfill &\cr
  \+\hfill    27&\hfill    10&\hfill 37.0\%&\hfill    14&\hfill 51.9\%&\hfill     2&\hfill  7.4\%&\cr
  \+\hfill   729&\hfill   288&\hfill 39.5\%&\hfill   256&\hfill 35.1\%&\hfill   184&\hfill 25.2\%&\cr
  \+\hfill 19683&\hfill  7158&\hfill 36.4\%&\hfill  6290&\hfill 32.0\%&\hfill  6234&\hfill 31.7\%&\cr
}$$

The right side of \figref{adjcorn} shows a cell with improved isotropy.  The
top and bottom halves have 5 vertical segments each; the left and
right halves have 4 lateral segments each; and the front and back
halves have 4 front-to-back segments each.  There are only 2 straight
runs:

$$\vbox{\settabs 9\columns
  \+\hfill nodes&\hfill   &$j=    0$\hfill &\hfill   &$j=    1$\hfill &\hfill   &$j=    2$\hfill &\cr
  \+\hfill    27&\hfill     8&\hfill 29.6\%&\hfill     8&\hfill 29.6\%&\hfill    10&\hfill 37.0\%&\cr
  \+\hfill   729&\hfill   248&\hfill 34.0\%&\hfill   240&\hfill 32.9\%&\hfill   240&\hfill 32.9\%&\cr
  \+\hfill 19683&\hfill  6552&\hfill 33.3\%&\hfill  6578&\hfill 33.4\%&\hfill  6552&\hfill 33.3\%&\cr
}$$

\beginsection{10 Filling All of Space}

So far the self-similarity of space-filling curves has been realized
by identifying the origin of the Hamiltonian-path with the origin of
the path of the first sub-cell, which results in ranges with
non-negative coordinates.  For serpentine diagonal cells (with odd $s$)
it is possible to identify the midpoint of the Hamiltonian-path with
the midpoint of the center sub-cell, extending the range to all integer
coordinates.  While the treatment above could be redeveloped using a
symmetric modulo operator, it is simpler to use scaled offsets of the
recurrence \eqref{R} inputs and outputs.

$$\eqalign{
        l(u) &= \left\lceil\log_{s^{d^2}}(2\cdot(1+|u|))\right\rceil\cr
        Q'(u) &= Q\left(u+\left\lfloor s^{d^2 l(u)}/2\right\rfloor\right)
        -\left(\left\lfloor s^{d\cdot l(u)}/2\right\rfloor,\left\lfloor s^{d\cdot l(u)}/2\right\rfloor,\dots\right) \cr} \eqdef{Q'}$$

$s^{d^2 l(u)}$ finds the smallest integer power of $s^{d^2}$ half of
which is large enough to offset $-|u|$ to be positive.  Half of that
is added to $u$ and the corresponding offset is subtracted from each
coordinate returned.  Because the base-$s$ digits of $s^n/2$ are
$\lfloor s/2\rfloor, \lfloor s/2\rfloor, \dots $, the offsets bias
$u=0$ to the center sub-cell at each scale.

With such auto-scaling $u$ can take any integer value and $Q'(u)$ can
reach any point with integer-coordinates.  How do these curves spread
to the whole plane?  \figref{center} shows $Q'(u)$ for $u$ over the range of
$-3280 = -\left\lfloor9^4/2\right\rfloor$ to $0$.  The Peano and its
nearly-isotropic variant wrap around the origin in a spiral; the
positive $u$ values do so also.  The isotropic variant wraps back and
forth around the origin.
\smallskip
\vbox{{\settabs 3\columns
\+\hfill\fig{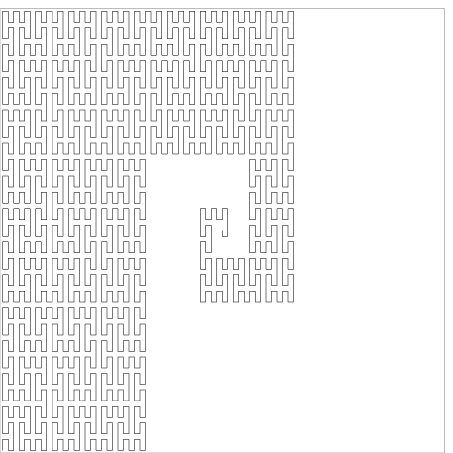}\hfill&
  \hfill\fig{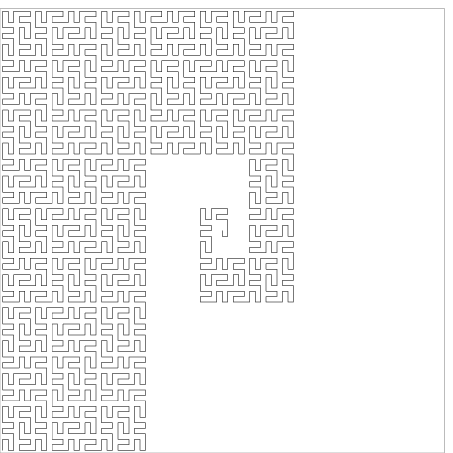}\hfill&
  \hfill\fig{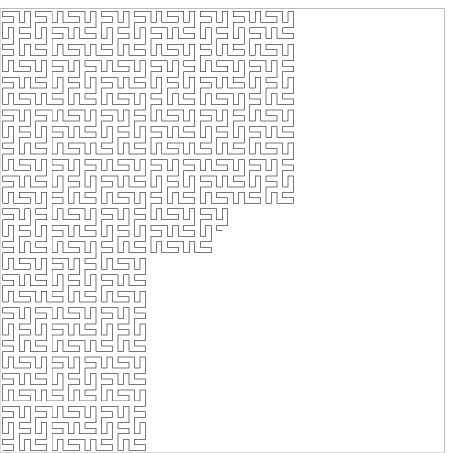}\hfill&\cr
\+\hfill Peano\hfill&
  \hfill Nearly Isotropic\hfill&
  \hfill Isotropic\hfill&\cr}
\centerline{\figdef{center}}}

The inverse function scales according to the largest coordinate in
input vector $V$.

$$\eqalign{ l'(V)
        &= \lceil\log_{s^d}\left(2\cdot\left(1+\max(|V_0|,\dots,|V_{d-1}|)\right)\right)\rceil\cr
        q'(V) &= q\left(V+\left(\left\lfloor s^{d\cdot
        l'(V)}/2\right\rfloor,\left\lfloor s^{d\cdot
        l'(V)}/2\right\rfloor,\dots\right)\right) -\left\lfloor s^{d^2
        l'(V)}/2\right\rfloor\cr} \eqdef{q'}$$

Equation \eqref{q'} $q'$ is particularly useful for reducing the dimension
of data-sets whose bounds are not known in advance.

Because the most significant digit of $F$ and $f$ (and $E$ and $e$) is
always the most significant fractional digit, centering the unit real
space-filling function is easy:

$$\eqalignno{
F'(y)&=F\left(y+{1\over 2}\right)-\left({1\over 2},{1\over 2},\dots\right)&\eqdef{F'}\cr
f'(Y)&=f\left(Y+\left({1\over 2},{1\over 2},\dots\right)\right)-{1\over 2}&\eqdef{f'}\cr
E'(y)&=E\left(y+{1\over 2}\right)-\left({1\over 2},{1\over 2},\dots\right)&\eqdef{E'}\cr
e'(Y)&=e\left(Y+\left({1\over 2},{1\over 2},\dots\right)\right)-{1\over 2}&\eqdef{e'}\cr
}$$

\beginsection{11 Dimension Reduction Performance}

For dimension reduction it is desirable that points that are close in
$d$-dimensional space map to scalar values that are close and that
points that are distant map to scalar values that are distant.

\figdefn{Hperf}
\figdefn{Pperf}
\figdefn{PNIperf}
\figdefn{PIperf}
\vbox{\settabs 2\columns
\+\hfill\fig{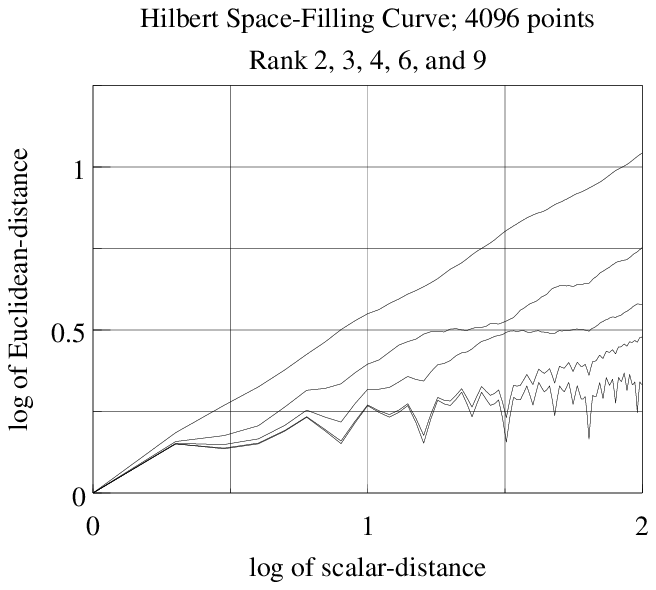}\hfill&
  \hfill\fig{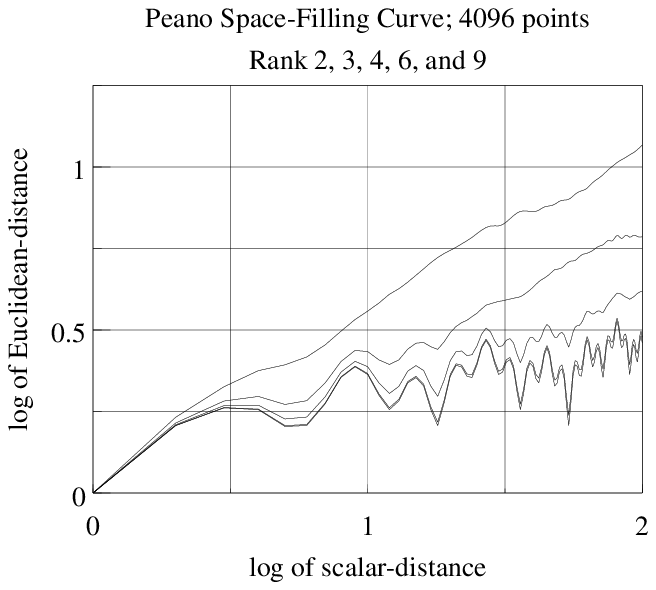}\hfill&\cr
\+\hfill \figref{Hperf}\hfill&
  \hfill \figref{Pperf}\hfill&\cr
\+\hfill\fig{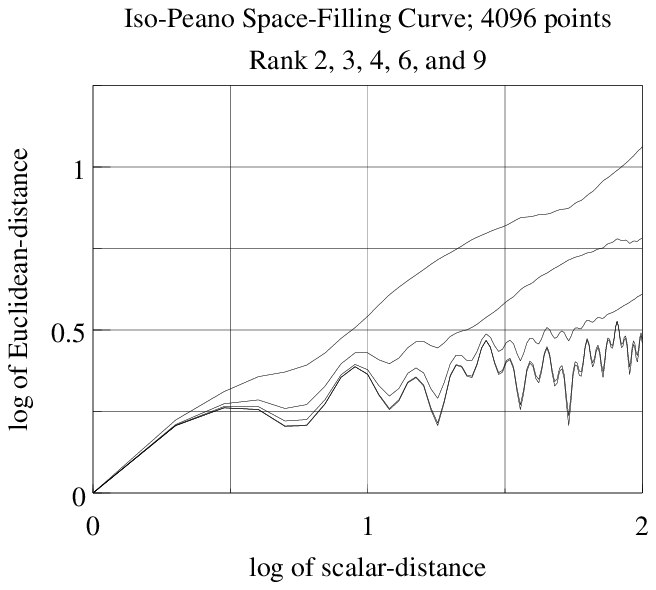}\hfill&
  \hfill\fig{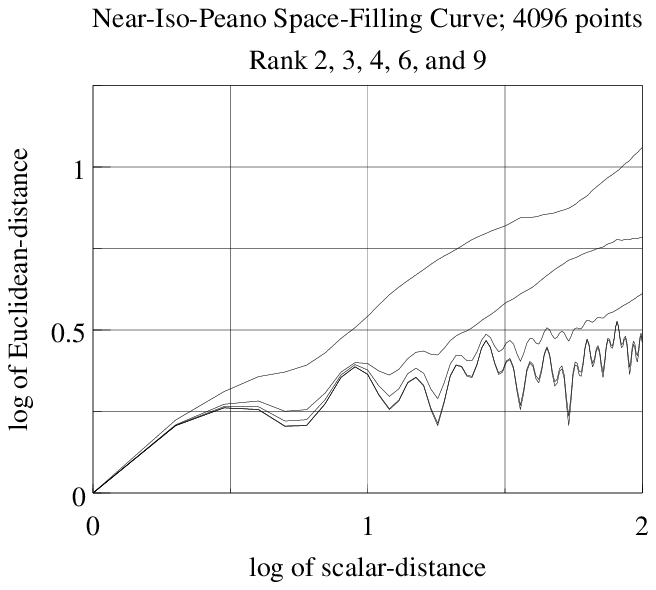}\hfill&\cr
\+\hfill \figref{PIperf}\hfill&
  \hfill \figref{PNIperf}\hfill&\cr
\+\hfill\fig{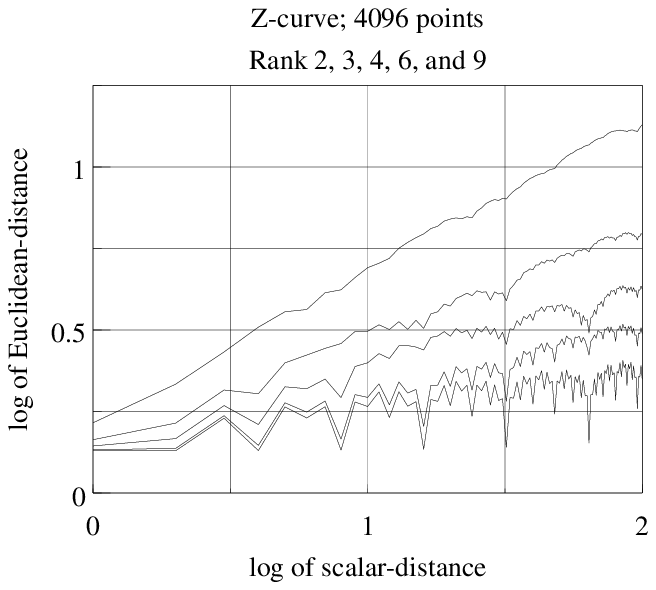}\hfill&
  \hfill\vbox{\fig{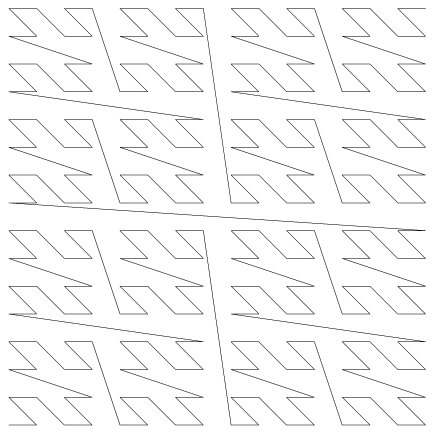}\bigskip{\hbox{Z-curve}}\medskip}\hfill&\cr
\+\hfill \figdef{Zperf}\hfill&
  \hfill \figdef{Z}\hfill&\cr}

\figrefs{Hperf} through \figrefn{Zperf} are log-log plots of the
average euclidean distance $||Q(v_1)-Q(v_2)||$ versus scalar lengths
$|v1-v2|$ from 1 to 100 for Hilbert ($s=2$), Peano-like ($s=3$), and Z
curves of ranks 2, 3, 4, 6, and~9.  The top trace of each plot is for
rank $d=2$; each successively lower curve is of higher rank.

\figref{Z} shows the Z-curve (which Sagan attributes to Lebesgue in
\cite{Sagan:1994}), one of the simplest space-filling
functions, although it isn't continuous.  It forms each vector
coordinate from every $d$-th bit of the scalar.  It is self-avoiding,
but contains some long edges.

The Z-curve has the widest spread and is particularly poor in
2-dimensions.  Worse still is that most unit length scalar edges map
to coordinates which are more than one unit separated.  Thus the
average displacement for unit edges averages above 1.6 for
2-dimensions and above 1.35 for the other ranks.

Ranks 6 and 9 appear coincident in the graphs of Peano-like curves.
Except for the Z-curve, the rank-2 curves show euclidean distance
being slightly longer than the square-root of scalar distance.  Curves
for the higher ranks are not as close to the $d$th root.

The great similarity between \figrefs{Pperf}, \figrefn{PNIperf},
and \figrefn{PIperf} indicate that these measurements are largely
unaffected by isotropy and run-length distribution.

\beginsection{12 Praxis}

The cell properties $H(t)$, $d$, $i$, $s$, $N(t)$, $X(t)$,
$H^{-1}(V)$, and $k(t)$ can be precomputed and stored within the
data-structure for the cell.  If the shift-count $\sigma(t)$ is also
precomputed:

$$\sigma(t)=\cases
        {i-k(t),& if $H(t)$ is an adjacent-corner cell;\cr
         0,     & if $H(t)$ is a Peano cell;\cr
         t,     & if $H(t)$ is a near-isotropic opposite-corner cell ($A'$);\cr
         t+1,   & if $H(t)$ is an isotropic opposite-corner cell ($A''$).\cr}
 \eqdef{sigma(t)}$$

then all the alignment functions can be computed using the same formula:

$$A_j(V,t,w) = \cases
              {V_{j+\sigma(t)\bmod d},      & if $N_j(t) = 0$;\cr
              w-V_{j+\sigma(t)\bmod d},     & otherwise.\cr} \eqdef{A_jsigma}$$

Similarly, the inverse alignment functions can computed using a single
formula:

$$A^{-1}_j(V,t,w)=\cases
              {V_{j-\sigma(t)\bmod d},      & if $N_j(t) = 0$;\cr
              w-V_{j-\sigma(t)\bmod d},     & otherwise.\cr} \eqdef{A^{-1}_jsigma}$$

Thus the recurrences treat diagonal-corner and adjacent-corner
patterns uniformly.

Equations \eqref{R}, \eqref{f}, \eqref{Q'}, and \eqref{q'} have terms
involving ceilings of logarithms.  While mathematically correct,
practical implementations will compute these quantities using integers
as suggested by their descriptions in the text.  Floored division
($\lfloor u/m\rfloor$) is simply integer division.

When $F(y)$ or $E'(y)$ is computed with inexact numbers, small
increments ($\approx1\times10^{-9}$) must be added to floor function
inputs so that rounding and truncation don't foil the calculation.

R5RS Scheme (Kelsey et al \cite{R5RS}) code implementing cell
construction and the $Q$, $q$, $Q'$, and $q'$ integer recurrences is
found in the SLIB Scheme Library's {\tt space-filling} package:\par
{\tt
http://cvs.savannah.gnu.org/viewvc/slib/slib/rmdsff.scm?view=markup}.
Documentation is at {\tt
http://people.csail.mit.edu/jaffer/slib/Multidimensional-Space\_002dFilling-Curves.html}

\beginsection{13 Conclusion}

Algorithms implementing the recurrences presented here compute the
space-filling coordinates or the scalar value (from space-filling
coordinates) with geometric convergence.  These algorithms work with
serpentine cells of any rank $d\ge2$ and side length $s\ge2$ and many
non-serpentine cells.

The serpentine cell algorithm produces cells for Hilbert and Peano
space-filling curves and functions, their pandimensional analogs and
inverses, as well as novel pandimensional space-filling curves with
improved properties.

By offsetting and scaling the scalar and coordinates, the functions
and curves can be extended beyond the unit (hyper)cube, even to all of
$d$-space for serpentine diagonal-travel cells (odd $s\ge3$).

The performance of space-filling curves for dimension-reduction was
measured for multidimensional Hilbert and Peano curves; Hilbert curves
performed somewhat better than Peano curves and their isotropic
variants.


\beginsection{References}

\bibliographystyle{unsrt}
\bibliography{RMDSFF}

\vfill\eject
\bye